\documentclass[12pt]{article}
\usepackage{hyperref,slashed,graphicx,color,amsmath,amssymb}

\begin{document}

\vskip 1cm
\begin{center}
{ \bf\large
A nondiagrammatic calculation of the $\rho$ parameter from heavy fermions
\vskip 0.2cm}
\vskip 0.3cm
{Hong-Hao Zhang\footnote{Email: zhh98@mail.sysu.edu.cn}}
\vskip 0.3cm
{\it \small School of Physics and Engineering, Sun Yat-Sen University, Guangzhou
510275, China}
\end{center}
\vspace{0.3cm}
\begin{center}
{\bf Abstract}
\end{center}

A simple nondiagrammatic evaluation of the nondecoupling effect of
heavy fermions on the Veltman's $\rho$ parameter is presented in
detail. This calculation is based on the path integral approach, the
electroweak chiral Lagrangian formalism, and the Schwinger proper
time method.

\newpage

Although the Standard Model (SM) of particle physics was in very
good agreement with the data during the past thirty years or so,
there are strong indications that it is just a low-energy effective
theory. For example, we do not discover the Higgs particle; we do
not know how to unify the SM with gravity, why the SM gauge group is
$SU(3)_c\times SU(2)_L\times U(1)_Y$, and why there are 3 families
of particles and so on; and the SM suffers the hierarchy problem, the
unnaturalness problem, etc. At present there are many new physics
extensions beyond the SM. Although we do not know whether nature
really behaves like one of them or not, we can estimate their
effects on the current electroweak precision observables. The
Veltman's $\rho$ parameter, defined by $m_W^2/(m_Z^2c_W^2)$, is of
particular interest among these observables \cite{Veltman:1977kh}.
In the SM the $\rho$ parameter is equal to 1 at tree level, which is
protected by the custodial $SU(2)$ symmetry in the Higgs sector of
the SM. This symmetry overlaps with the local $SU(2)_L$ gauge
symmetry. Mass splittings in weak isospin doublets break this
symmetry. Their effect leaks into radiative corrections which make
$\rho$ differ slightly from 1. In model-independent studies the
$\rho$ parameter is related to the $T$ parameter in the
Peskin-Takeuchi's formalism \cite{Peskin:1990zt,Peskin:1991sw}, and
is also related to the $\beta_1$ parameter in the electroweak chiral
Lagrangian
\cite{Appelquist:1993ka,Wang:2001xz,Zhang:2007sm,Zhang:2007zi,Lang:2008ap,
Lang:2009we,Zhang:2009vf}. Whether the $\rho$ parameter is 1 or not
is an indication whether the custodial symmetry is preserved or not
in a generic theory. Current electroweak data fitting favors the
$\rho$ parameter being close to 1, that is, $1.00989\leq \rho^{\rm
exp}\leq 1.01026$ \cite{Yao:2006px,Jung:2007tr}, which puts
stringent constraints on many new physics models. In literature the
nondecoupling effects of a heavy fermion weak doublet or even an
$n$-plet on the $\rho$ parameter were already obtained using the
Feynman diagrammatic calculation
\cite{Peskin:1991sw,Appelquist:1993ka,Zhang:2006de}. In the present
study we give a new calculational method of the contributions of
heavy fermions to the $\rho$ parameter without referring to Feynman
diagrams. The calculation itself is quite simple in the present
method, although one needs some basic knowledge of the electroweak
chiral Lagrangian and the Schwinger proper time method
\cite{Schwinger:1951nm,DeWitt:1965jb,Lu:2002fx,Lu:2002en}.

We start by considering a weak doublet of heavy fermions $U$ and
$D$, with masses $m_U$ and $m_D$ respectively, and representations
of $SU(2)_L\times U(1)_Y$ as
\begin{eqnarray}
\psi_L\equiv\begin{pmatrix}U\\D\end{pmatrix}_L \sim (2,0)\;,\qquad
U_R \sim (1,\frac{1}{2})\;,\qquad D_R \sim (1,-\frac{1}{2})
\end{eqnarray}
At some energy scale higher than the electroweak scale, the
effective action of the heavy fermion sector can be written in a
chiral invariant form as
\begin{eqnarray}
S_{\rm eff}[U,W_\mu^a,B_\mu,\bar{\psi},\psi]&=&\int
d^4x\bigg[\bar{\psi}_L\big(i\slashed{\partial}-g_2\frac{\tau^a}{2}\slashed{W}^a\big)\psi_L
+\bar{\psi}_R\big(i\slashed{\partial}
-g_1\frac{\tau^3}{2}\slashed{B}\big)\psi_R\nonumber\\
&&-(\bar{\psi}_LUM\psi_R+\bar{\psi}_RMU^\dag\psi_L)\bigg]~~~~~\label{action-fm-1}
\end{eqnarray}
where $\tau^a$($a=1,2,3$) are Pauli matrices; $g_1$ and $g_2$
($B_\mu$ and $W_\mu^a$) are the $U(1)_Y$ and $SU(2)_L$ gauge
couplings (fields), respectively; the dimensionless unitary
unimodular matrix $U(x)$ is the nonlinear realization of the
Goldstone boson fields in the electroweak chiral Lagrangian; and the
fermion mass matrix $M\equiv\mathrm{diag}(m_U,~m_D)$. The last two
terms of eq.\eqref{action-fm-1} will restore the usual fermion mass
terms when the physical gauge, i.e. the unitary gauge $U=1$, is
taken. The $U(x)$ field represents the degree of freedom of the
would-be Goldstone bosons from the electroweak symmetry breaking,
and its transformation under $SU(2)_L\times U(1)_Y$ is given by
\begin{eqnarray}
U(x)\rightarrow V_L(x)U(x)V_R^\dag(x)
\end{eqnarray}
where $V_L(x)=\exp\{i\frac{\tau^a}{2}\theta^a(x)\}$ and
$V_R(x)=\exp\{i\frac{\tau^3}{2}\theta^0(x)\}$ with $\theta^a(x)$ and
$\theta^0(x)$ being the $SU(2)_L$ and $U(1)_Y$ group parameters,
respectively. To derive the contribution of the heavy fermions to
the $\rho$ parameter, or equivalently the $\beta_1$ parameter in the
electroweak chiral Lagrangian, we need to integrate out the heavy
fermions above the electroweak scale which can be formulated as
\begin{eqnarray}
\exp(iS_{\rm
EW}[U,W_\mu^a,B_\mu])=\int\mathcal{D}\bar{\psi}\mathcal{D}\psi\exp(iS_{\rm
eff}[U,W_\mu^a,B_\mu,\bar{\psi},\psi])\label{stra}
\end{eqnarray}
where $S_{\rm EW}$ denotes the nondecoupling contribution of heavy
fermions to the effective action just at the electroweak scale.
Following Ref.\cite{Zhang:2007zi}, we can make the chiral
decomposition of the $U(x)$ field as follows:
\begin{eqnarray}
U(x)=\xi_L^\dag(x)\xi_R(x)
\end{eqnarray}
where $\xi_L(x)=\exp\{i\frac{\tau^a}{2}\phi^a(x)\}$ and
$\xi_R(x)=\exp\{i\frac{\tau^3}{2}\phi^0(x)\}$, and their
transformation under $SU(2)_L\times U(1)_Y$ are
\begin{eqnarray}
\xi_L(x)\rightarrow h(x)\xi_L(x)V_L^\dag(x)\;,\qquad
\xi_R(x)\rightarrow h(x)\xi_R(x)V_R^\dag(x)\label{trans-xi}
\end{eqnarray}
where $h(x)=\exp\{i\frac{\tau^3}{2}\theta_h(x)\}$ belongs to an
induced hidden local $U(1)$ symmetry group. Now we make a special
$SU(2)_L\times U(1)_Y$ chiral rotation as: $V_L(x)=\xi_L(x)$,
$V_R(x)=\xi_R(x)$, under which $U(x)$ is rotated to be 1 and the
explicit $U$-dependence of $S_{\rm eff}$ disappears. Thus
eq.\eqref{action-fm-1} becomes
\begin{eqnarray}
S_{\rm eff}[V_\mu^a,V_\mu^0,\bar{\psi}_\xi,\psi_\xi]=\int
d^4x\bar{\psi}_{\xi}(i\slashed{\partial}+\slashed{v}+\slashed{a}\gamma_5-s-m)\psi_{\xi}
\end{eqnarray}
where $m\equiv(m_U+m_D)/2$, the traceless matrix $s\equiv\Delta
m\cdot\tau^3$ with $\Delta m\equiv(m_U-m_D)/2$, and the rotated
fields are defined as
\begin{eqnarray}
&&\psi_{\xi}(x)=P_L\xi_L(x)\psi_L(x)+P_R\xi_R(x)\psi_R(x)\\
&&v_\mu(x)\equiv-\frac{1}{2}[g_2\frac{\tau^a}{2}V_\mu^a(x)+g_1\frac{\tau^3}{2}V_\mu^0(x)]\\
&&a_\mu(x)\equiv\frac{1}{2}[g_2\frac{\tau^a}{2}V_\mu^a(x)-g_1\frac{\tau^3}{2}V_\mu^0(x)]\\
&&\mbox{with}\qquad g_2\frac{\tau^a}{2}V_\mu^a(x)\equiv
\xi_L[g_2\frac{\tau^a}{2}W^a_\mu(x)-i\partial_\mu]\xi_L^\dag\label{def-rot-gaug-1}\\
&&\qquad \qquad g_1\frac{\tau^3}{2}V_\mu^0(x)\equiv
\xi_R[g_1\frac{\tau^3}{2}B_\mu(x)-i\partial_\mu]\xi_R^\dag\label{def-rot-gaug-2}
\end{eqnarray}
According to eq.\eqref{trans-xi}, the transformations of the above
rotated fields under $SU(2)_L\times U(1)_Y$ are given by
\begin{eqnarray}
&&\psi_{\xi}(x)\rightarrow h(x)\psi_{\xi}(x)\\
&&g_2\frac{\tau^a}{2}V_\mu^a \rightarrow
h(x)[g_2\frac{\tau^a}{2}V_\mu^a-i\partial_\mu]h^\dag(x)\\
&&g_1\frac{\tau^3}{2}V_\mu^0 \rightarrow
h(x)[g_1\frac{\tau^3}{2}V_\mu^0-i\partial_\mu]h^\dag(x)
\end{eqnarray}
Thus, the chiral symmetry $SU(2)_L\times U(1)_Y$ covariance of the
unrotated fields has been transferred totally to the hidden symmetry
$U(1)$ covariance of the rotated fields. We further find that
$a_\mu(x)$ transforms covariantly: $a_\mu(x)\to
h(x)a_\mu(x)h^\dag(x)$, while $v_\mu(x)$ transforms as the ``gauge
field" of the hidden local $U(1)$ symmetry: $v_\mu(x)\to
h(x)[v_\mu(x)+i\partial_\mu]h^\dag(x)$. Accordingly, for any
operator $O(x)$ which transforms covariantly under the hidden local
symmetry: $O(x)\to h(x)O(x)h^\dag(x)$, we can define its covariant
derivative as: $d_\mu O(x)\equiv \partial_\mu
O(x)-i[v_\mu(x),O(x)]$. It is important to note that the traces of
some combinations of the rotated fields can be related to the terms
of the electroweak chiral Lagrangian. First of all, from
eqs.\eqref{def-rot-gaug-1} and \eqref{def-rot-gaug-2} it is straightforward to obtain:
\begin{eqnarray}
g_2\frac{\tau^a}{2}V_\mu^a(x)-g_1\frac{\tau^3}{2}V_\mu^0(x)=-i\xi_RX_\mu\xi_R^\dag
\label{eq-12}
\end{eqnarray}
where $X_\mu\equiv U^\dag(D_\mu U)$. Thus, the axial vector field
$a_\mu=-\frac{i}{2}\xi_R X_\mu\xi_R^\dag$. And we can also write $s$
in this covariant form: $s=\xi_R \Delta m\cdot\tau^3\xi_R^\dag$,
since $\tau^3$ commutes with $\xi_R$ and $\xi_R^\dag$. On the other
hand, we have
\begin{eqnarray}
v_\mu=\frac{i}{2}\xi_R
X_\mu\xi_R^\dag-\xi_Rg_1\frac{\tau^3}{2}B_\mu\xi_R^\dag+i\xi_R(\partial_\mu\xi_R^\dag)
\end{eqnarray}
which leads to the following two relations:
\begin{eqnarray}
i\xi_R^\dag v_\mu&=&-\frac{1}{2}
X_\mu\xi_R^\dag-ig_1\frac{\tau^3}{2}B_\mu\xi_R^\dag-(\partial_\mu\xi_R^\dag)\\
-iv_\mu\xi_R&=&\frac{1}{2}\xi_R
X_\mu+i\xi_Rg_1\frac{\tau^3}{2}B_\mu-(\partial_\mu\xi_R)
\end{eqnarray}
i.e.
\begin{eqnarray}
&&(\partial_\mu\xi_R^\dag)+i\xi_R^\dag v_\mu=(-\frac{1}{2}
X_\mu-ig_1\frac{\tau^3}{2}B_\mu)\xi_R^\dag\\
&&(\partial_\mu\xi_R)-iv_\mu\xi_R=\xi_R(\frac{1}{2}
X_\mu+ig_1\frac{\tau^3}{2}B_\mu)
\end{eqnarray}
Thus, for any chiral rotated field $f\equiv \xi_R F\xi_R^\dag$, we
have
\begin{eqnarray}
d_\mu f&\equiv&\partial_\mu f-i[v_\mu,f]\nonumber\\
&=&\xi_R(\partial_\mu F)\xi_R^\dag+\xi_R(\frac{1}{2}
X_\mu+ig_1\frac{\tau^3}{2}B_\mu)F\xi_R^\dag+\xi_RF(-\frac{1}{2}
X_\mu-ig_1\frac{\tau^3}{2}B_\mu)\xi_R^\dag\nonumber\\
&=&\xi_R\bigg\{(D_\mu F)+\frac{1}{2}[X_\mu,F]\bigg\}\xi_R^\dag
\end{eqnarray}
where $D_\mu F\equiv\partial_\mu F+[ig_1\frac{\tau^3}{2}B_\mu,F]$.
In particular, if $f=s=\xi_R \Delta m\cdot\tau^3\xi_R^\dag$, then
$F=\Delta m\cdot\tau^3$ and we have
\begin{eqnarray}
d_\mu s=\xi_R\bigg\{D_\mu (\Delta
m\cdot\tau^3)+\frac{1}{2}[X_\mu,\Delta
m\cdot\tau^3]\bigg\}\xi_R^\dag =\frac{1}{2}\Delta
m\xi_R[X_\mu,\tau^3]\xi_R^\dag
\end{eqnarray}
and hence,
\begin{eqnarray}
d_\mu(d^\mu s)=\xi_R\bigg\{\frac{1}{2}\Delta m[(D_\mu
X^\mu),\tau^3]+\frac{1}{4}\Delta
m\big[X_\mu,[X^\mu,\tau^3]\big]\bigg\}\xi_R^\dag
\end{eqnarray}
Using the above relations, we can obtain the following identities:
\begin{eqnarray}
&&{\rm tr}(sa_\mu sa^\mu)=-\frac{1}{4}\Delta m^2{\rm
tr}(\tau^3X_\mu){\rm tr}(\tau^3X^\mu)+\frac{1}{4}\Delta m^2{\rm
tr}(X_\mu X^\mu)\label{eq-21}\\
&&{\rm tr}[(d_\mu s)(d^\mu s)]=-{\rm tr}[sd_\mu(d^\mu
s)]\nonumber\\
&&\qquad\qquad\qquad=\frac{1}{2}\Delta m^2{\rm tr}(\tau^3X_\mu){\rm
tr}(\tau^3X^\mu)-\Delta m^2{\rm tr}(X_\mu X^\mu)\label{eq-22}
\end{eqnarray}
which are relevant to the $\beta_1$ term in the electroweak chiral
Lagrangian. Now, we proceed to integrate out the heavy fermion
fields to get their contributions to the low-energy electroweak
effective action. The integral measure of the heavy fermions remains
unchanged under the special chiral rotation, since these fermions
have an anomaly-free assignment of gauge charges. Thus,
eq.\eqref{stra} can be written as
\begin{eqnarray}
&&\hspace{-0.7cm}\exp(iS_{\rm
EW}[U,W_\mu^a,B_\mu])\nonumber\\
&=&\int\mathcal{D}\bar{\psi}_\xi\mathcal{D}\psi_\xi\exp\bigg[i\int
d^4x\bar{\psi}_{\xi}(i\slashed{\partial}
+\slashed{v}+\slashed{a}\gamma_5-s-m)\psi_{\xi}\bigg]_M\nonumber\\
&=&\int\mathcal{D}\bar{\psi}_\xi\mathcal{D}\psi_\xi \exp\bigg[-\int
d^4x\bar{\psi}_{\xi}(\slashed{\partial}-i\slashed{v}-i\slashed{a}\gamma_5-s+m)\psi_{\xi}
\bigg]_E\label{stra-1}
\end{eqnarray}
where the subscripts $M$ and $E$ indicate that the expressions are
written respectively in the Minkowski spacetime and in the Euclidean
spacetime (See Appendix \ref{app-relation-Min-Eu} for the conversion
relations of quantities in these two spaces). Here and henceforth,
we mostly work in the Euclidean spacetime and, for convenience, we
drop the subscript $E$ in the expressions until further notice.
Eq.\eqref{stra-1} gives
\begin{eqnarray}
iS_{\rm
EW}[U,W_\mu^a,B_\mu]=\ln\mathrm{Det}(D+m)=\mathrm{Tr}\ln(D+m)
\end{eqnarray}
where $D\equiv
\slashed{\partial}-i\slashed{v}-i\slashed{a}\gamma_5-s$. Since the
imaginary part of the fermion determinant corresponds to the
Wess-Zumino-Witten anomaly term
\cite{Wess:1971yu,Witten:1983tw,Ma:2003uv}, for any anomaly-free
underlying model we only need to consider the calculation of the
real part of the fermion determinant. Using the Schwinger proper
time formula \cite{Schwinger:1951nm}, we have
\begin{eqnarray}
{\rm Re \; Tr}\ln(D+m)&=&\frac{1}{2}{\rm Tr}\ln[(D^\dag+m)(D+m)]\nonumber\\
&=&-\frac{1}{2}\lim_{\Lambda\rightarrow\infty}\int
d^4x\int_{\frac{1}{\Lambda^2}}^\infty\frac{d\tau}{\tau}{\rm
tr}_{c,f,l}\langle x|e^{-\tau(D^\dag+m)(D+m)}|x\rangle
\label{schwingerpropertime}
\end{eqnarray}
where ${\rm tr}_{c,f,l}$ denotes taking trace with respect to color
(or technicolor), flavor, and Lorentz indices, and the operator
$(D^\dag+m)(D+m)$ in the exponential can be further simplified as
follows:
\begin{eqnarray}
(D^\dag+m)(D+m)=E-\nabla^2+m^2
\end{eqnarray}
with
\begin{eqnarray}
&&E\equiv-2ms-2ima\!\!\!/\,\gamma_5+\frac{i}{4}[\gamma^\mu,\gamma^\nu]R_{\mu\nu}+\gamma^\mu
d_\mu s+i\gamma^\mu\{s,a_\mu\}\gamma_5+s^2\\
&&\nabla_\mu\equiv\partial_\mu-iv_\mu-ia_\mu\gamma_5\\
&&R_{\mu\nu}\equiv i[\nabla_\mu,\nabla_\nu]=V_{\mu\nu}+(d_\mu
a_\nu-d_\nu
a_\mu)\gamma_5-i[a_\mu,a_\nu]\\
&&V_{\mu\nu}\equiv i[\partial_\mu-iv_\mu,
\partial_\nu-iv_\nu]=\partial_\mu v_\nu-\partial_\nu
v_\mu-i[v_\mu,v_\nu]\\
&&d_\mu s\equiv\partial_\mu s-i[v_\mu,s]
\end{eqnarray}
Thus eq.\eqref{schwingerpropertime} can be written as:
\begin{eqnarray}
{\rm Re \;
Tr}\ln(D+m)=-\frac{1}{2}\lim_{\Lambda\rightarrow\infty}\int
d^4x\int_{\frac{1}{\Lambda^2}}^\infty\frac{d\tau}{\tau}e^{-\tau
m^2}{\rm tr}_{c,f,l}\langle
x|e^{-\tau(E-\nabla^2)}|x\rangle\label{schwingerpropertime-2}
\end{eqnarray}
where $\langle x|e^{-\tau(E-\nabla^2)}|x\rangle$ is the so-called
heat kernel, and it can be expanded in powers of $\tau$ (and in
powers of momenta as well) using the Seely-DeWitt expansion formula
\cite{DeWitt:1965jb} as
\begin{eqnarray}
\langle
x|e^{-\tau(E-\nabla^2)}|x\rangle&=&\frac{1}{16\pi^2}\bigg\{\frac{1}{\tau^2}-\frac{E}{\tau}+\big(\frac{1}{2}E^2-\frac{1}{6}[\nabla_\mu,[\nabla^\mu,E]]
-\frac{1}{12}R_{\mu\nu}R^{\mu\nu}\big)\nonumber\\
&&+\tau\big(-\frac{1}{6}E^3+\frac{1}{12}([\nabla_\mu,[\nabla^\mu,E]]E+E[\nabla_\mu,[\nabla^\mu,E]]\nonumber\\
&&+[\nabla_\mu,E][\nabla^\mu,E])\big)+\frac{\tau^2}{24}E^4\bigg\}+\mathcal{O}(\tau^3)\label{eq-29}
\end{eqnarray}
From the structure of $E$ and $\nabla_\mu$, we find that only the
traces of the following terms in eq.\eqref{eq-29} have contributions
to the coefficient of ${\rm tr}(sa_\mu sa^\mu)$:
\begin{eqnarray}
&&\mathrm{tr}_l\bigg(\frac{1}{2}E^2\bigg)\ni 4sa_\mu sa^\mu\label{eq-30}\\
&&\mathrm{tr}_l\bigg(-\frac{\tau}{6}E^3\bigg)\ni -16\tau m^2sa_\mu sa^\mu\\
&&\mathrm{tr}_l\bigg(\frac{\tau}{12}\big([\nabla_\mu,[\nabla^\mu,E]]E+E[\nabla_\mu,[\nabla^\mu,E]]
+[\nabla_\mu,E][\nabla^\mu,E]\big)\bigg)\ni \frac{8}{3}\tau m^2sa_\mu sa^\mu\nonumber\\
\\
&&\mathrm{tr}_l\bigg(\frac{\tau^2}{24}E^4\bigg)\ni
\frac{16}{3}\tau^2m^4sa_\mu sa^\mu\label{eq-31}
\end{eqnarray}
where $\mathrm{tr}_l$ stands for taking trace with respect to
Lorentz spinor indices. Combing eqs (\ref{eq-30}-\ref{eq-31}), we
obtain
\begin{eqnarray}
\mathrm{tr}_l\langle x|e^{-\tau(E-\nabla^2)}|x\rangle\ni
\frac{1}{16\pi^2}(4-\frac{40}{3}\tau
m^2+\frac{16}{3}\tau^2m^4)sa_\mu sa^\mu
\end{eqnarray}
which, by subsequently taking trace with respect to color (or
technicolor) and flavor indices, leads to
\begin{eqnarray}
\mathrm{tr}_{c,f,l}\langle x|e^{-\tau(E-\nabla^2)}|x\rangle\ni
\frac{N_c}{16\pi^2}(4-\frac{40}{3}\tau
m^2+\frac{16}{3}\tau^2m^4)\mathrm{tr}_f(sa_\mu
sa^\mu)\label{eq-sasa}
\end{eqnarray}
where $N_c$ is the color (or technicolor) degree of freedom of the
heavy fermions under consideration. From eq.\eqref{eq-22} we see
that $\mathrm{tr}[(d_\mu s)(d^\mu s)]$ and $\mathrm{tr}[sd_\mu(d^\mu
s)]$ differ only by opposite sign, and thus we may recognize them as
just one independent term. Likewise, by analyzing the structure of
$E$ and $\nabla_\mu$, we find that only the traces of the following
terms in eq.\eqref{eq-29} give contributions to the coefficient of
$\mathrm{tr}[(d_\mu s)(d^\mu s)]$:
\begin{eqnarray}
&&\mathrm{tr}_l\bigg(\frac{1}{2}E^2\bigg)\ni 2(d_\mu s)(d^\mu s)\\
&&\mathrm{tr}_l\bigg(\frac{\tau}{12}\big([\nabla_\mu,[\nabla^\mu,E]]E+E[\nabla_\mu,[\nabla^\mu,E]]
+[\nabla_\mu,E][\nabla^\mu,E]\big)\bigg)\ni -\frac{4}{3}\tau
m^2(d_\mu s)(d^\mu s)\nonumber\\
\end{eqnarray}
Combing the above equations gives
\begin{eqnarray}
\mathrm{tr}_l\langle x|e^{-\tau(E-\nabla^2)}|x\rangle\ni
\frac{1}{16\pi^2}(2-\frac{4}{3}\tau m^2)(d_\mu s)(d^\mu s)
\end{eqnarray}
which further implies
\begin{eqnarray}
\mathrm{tr}_{c,f,l}\langle x|e^{-\tau(E-\nabla^2)}|x\rangle\ni
\frac{N_c}{16\pi^2}(2-\frac{4}{3}\tau m^2)\mathrm{tr}_f[(d_\mu
s)(d^\mu s)]\label{eq-dsds}
\end{eqnarray}
Now, taking eqs.\eqref{eq-sasa} and \eqref{eq-dsds} into account, we
finally arrive at:
\begin{eqnarray}
\mathrm{tr}_{c,f,l}\langle x|e^{-\tau(E-\nabla^2)}|x\rangle &\ni&
\frac{N_c}{16\pi^2}\bigg\{(4-\frac{40}{3}\tau
m^2+\frac{16}{3}\tau^2m^4)\mathrm{tr}_f(sa_\mu sa^\mu)\nonumber\\
&&+(2-\frac{4}{3}\tau m^2)\mathrm{tr}_f[(d_\mu s)(d^\mu s)]
\bigg\}\label{eq-combine-sasa-dsda}
\end{eqnarray}
In the following we will re-use the subscripts $E$ and $M$ to stand
for, respectively, the quantities in the Euclidean space and those
in the Minkowski space, and we will use $\mathrm{tr}$ as a short for
$\mathrm{tr}_f$, i.e. taking trace over flavor space. Substituting
eq.\eqref{eq-combine-sasa-dsda} into
eq.\eqref{schwingerpropertime-2}, we obtain
\begin{eqnarray}
&&\hspace{-0.7cm}{\rm Re \; Tr}\ln(D+m)\nonumber\\
&\ni& -\frac{1}{2}\int
d^4x_E\lim_{\Lambda\rightarrow\infty}\int_{\frac{1}{\Lambda^2}}^\infty
\frac{d\tau}{\tau}e^{-\tau
m^2}\frac{N_c}{16\pi^2}\bigg\{(4-\frac{40}{3}\tau
m^2+\frac{16}{3}\tau^2m^4)\mathrm{tr}(sa_\mu sa^\mu)\nonumber\\
&&+(2-\frac{4}{3}\tau m^2)\mathrm{tr}[(d_\mu s)(d^\mu s)]
\bigg\}_E\nonumber\\
&=&-\frac{1}{2}\frac{N_c}{16\pi^2}\int
d^4x_E\lim_{\Lambda\rightarrow\infty}\bigg\{
[4(-\gamma-\ln\frac{m^2}{\Lambda^2})-8]\mathrm{tr}(sa_\mu sa^\mu)\nonumber\\
&&+[2(-\gamma-\ln\frac{m^2}{\Lambda^2})-\frac{4}{3}]\mathrm{tr}[(d_\mu
s)(d^\mu s)]\bigg\}_E\nonumber\\
&=&\frac{i}{2}\frac{N_c}{16\pi^2}\int
d^4x_M\lim_{\Lambda\rightarrow\infty}\bigg\{
[4(-\gamma-\ln\frac{m^2}{\Lambda^2})-8]\mathrm{tr}(sa_\mu sa^\mu)\nonumber\\
&&+[2(-\gamma-\ln\frac{m^2}{\Lambda^2})-\frac{4}{3}]\mathrm{tr}[(d_\mu
s)(d^\mu s)]\bigg\}_M\label{schwingerpropertime-3}
\end{eqnarray}
where we have used
eqs.(\ref{eq-b1}-\ref{eq-b3}),\eqref{eq-a-sasa},\eqref{eq-a-dsds}
and $\int d^4x_E=i\int d^4x_M$. Together with eqs.\eqref{eq-21} and
\eqref{eq-22}, eq.\eqref{schwingerpropertime-3} gives
\begin{eqnarray}
&&\hspace{-0.7cm}{\rm Re \; Tr}\ln(D+m)\nonumber\\
&\ni&\frac{i}{2}\frac{N_c}{16\pi^2}\int
d^4x_M\lim_{\Lambda\rightarrow\infty}\bigg\{
[4(-\gamma-\ln\frac{m^2}{\Lambda^2})-8][-\frac{1}{4}\Delta m^2{\rm
tr}(\tau^3X_\mu){\rm tr}(\tau^3X^\mu)]\nonumber\\
&&+[2(-\gamma-\ln\frac{m^2}{\Lambda^2})-\frac{4}{3}]\frac{1}{2}\Delta
m^2{\rm tr}(\tau^3X_\mu){\rm tr}(\tau^3X^\mu)\bigg\}\nonumber\\
&=&i\frac{N_c}{24\pi^2}\Delta m^2\int d^4x_M{\rm
tr}(\tau^3X_\mu){\rm tr}(\tau^3X^\mu)\label{schwingerpropertime-4}
\end{eqnarray}
where, as expected, the divergence term
$(-\gamma-\ln\frac{m^2}{\Lambda^2})$ has been exactly canceled by
the two parts. Thus, eq.\eqref{schwingerpropertime-4} gives the
nondecoupling contribution of the heavy fermions to the electroweak
effective action. Comparing this term with the $\beta_1$ term of the
standard electroweak chiral Lagrangian leads to,
\begin{eqnarray}
{\cal L}_0^\prime=\frac{1}{4}\beta_1f^2[{\rm tr}(\tau^3
X_\mu)]^2=\frac{N_c}{24\pi^2}\Delta m^2[{\rm tr}(\tau^3 X_\mu)]^2
\end{eqnarray}
which, together with the well-known relations,
$4/f^2=e^2/(s^2c^2M_Z^2)$ and $\Delta m=(m_U-m_D)/2$, implies,
\begin{eqnarray}
\beta_1=\frac{N_c}{96\pi^2}\frac{e^2}{s^2c^2}\frac{(m_U-m_D)^2}{M_Z^2}\label{result-1}
\end{eqnarray}
where $c\equiv \cos\theta_W$, $s\equiv\sin\theta_W$. Since
$\rho-1=\alpha~T=2\beta_1$, we finally obtain
\begin{eqnarray}
\rho=1+\frac{N_c}{48\pi^2}\frac{e^2}{s^2c^2}\frac{(m_U-m_D)^2}{M_Z^2}\label{result-2}
\end{eqnarray}
Eqs.\eqref{result-1} and \eqref{result-2} exactly coincide with the
Feynman-diagram calculated results in previous works
\cite{Peskin:1991sw,Appelquist:1993ka,Zhang:2006de}.

In summary, we have presented a new nondiagrammatic calculation
method of the nondecoupling effect of heavy fermions on the $\rho$
parameter. As we have seen, a crucial step of the present method is
figuring out the coefficients of ${\rm tr}(sa_\mu sa^\mu)$ and ${\rm
tr}[(d_\mu s)(d^\mu s)]$ in ${\rm Re \, Tr}\ln(D+m)$. In the
ordinary power counting of the QCD chiral Lagrangian, the scalar
source term $s$ is regarded as of order $p^2$, and these two trace
terms are of order $p^6$ and their coefficients were rarely
considered in literature. However, in the power counting of the
present case, $s$ is just a constant, i.e. the mass splitting of the
heavy fermions, and thus these two trace terms are of order $p^2$
and must give contributions to the $\beta_1$ term in the electroweak
chiral Lagrangian. Our method is also applicable to the case of
several heavy fermion doublets with mass mixing, and the generalization of this
calculation would be straightforward.

\appendix
\section{Relations Between Minkowski Space and Euclidean Space \label{app-relation-Min-Eu}}
To fix the notation, in this appendix we briefly review the relations between
quantities in the Minkowski space and those in the Euclidean space.
We use a mostly minus metric for the Minkowski space,
$(g_{\mu\nu})_M=\mathrm{diag}(1,-1,-1,-1)$, and we use a positive
metric for the Euclidean space,
$(g_{\mu\nu})_E=\mathrm{diag}(1,1,1,1)$. The coordinates in these
two space are related by:
\begin{eqnarray}
x_M^0=-ix_E^0\;,\qquad x_M^i=x_E^i \quad(i=1,2,3)
\end{eqnarray}
which gives $\int d^4x_M=-i\int d^4x_E$. From the correspondence
$p_\mu \sim
\partial/(\partial x^\mu)$, we accordingly obtain the conversion
relation of the momenta: $(p_0)_M=i(p_0)_E$, $(p_i)_M=(p_i)_E$
($i=1,2,3$), which by raising the indices gives,
\begin{eqnarray}
(p^0)_M=i(p^0)_E\;,\qquad (p^i)_M=-(p^i)_E \quad
(i=1,2,3)\label{eq-a2}
\end{eqnarray}
The vector source field $v^\mu(x)$ and the axial vector source field
$a^\mu(x)$ are ordered to be converted exactly the same way as
$p^\mu$. Furthermore, a minus sign appears in the relation of the
inner products of two vectors in these two spaces, $(g_{\mu\nu}p^\mu
p^\nu)_M=(p^0p^0-p^ip^i)_M=(-p^0p^0-p^ip^i)_E=-(g_{\mu\nu}p^\mu
p^\nu)_E$, which suggests us to let the scalar source field $s(x)$
be converted by: $s_M=-s_E$. From these relations, we have
\begin{eqnarray}
&&{\rm tr}(sa_\mu sa^\mu)\bigg|_M=-{\rm tr}(sa_\mu sa^\mu)\bigg|_E\label{eq-a-sasa}\\
&&{\rm tr}[(d_\mu s)(d^\mu s)]\bigg|_M=-{\rm tr}[(d_\mu s)(d^\mu
s)]\bigg|_E\label{eq-a-dsds}
\end{eqnarray}
On the other hand, since the Dirac matrices satisfy the Clifford
algebra, $\{\gamma^\mu, \gamma^\nu\}=2g^{\mu\nu}$, with $g^{\mu\nu}$
different in these two spaces, their conversion relations are given
by
\begin{eqnarray}
(\gamma^0)_M=(\gamma^0)_E\;,\qquad (\gamma^k)_M=i(\gamma^k)_E \quad
(k=1,2,3)\label{eq-a4}
\end{eqnarray}
In the Minkowski space, $(\gamma^0)_M$ is set to be hermitian, while
the other 3 Dirac matrices are anti-hermitian. And thus
$(\gamma_5\equiv i\gamma^0\gamma^1\gamma^2\gamma^3)_M$ is hermitian.
As a result, in the Euclidean space all the Dirac matrices
$(\gamma^\mu)_E$ are hermitian, so is
$(\gamma_5\equiv\gamma^0\gamma^1\gamma^2\gamma^3)_E$. Moreover, it
is easy to check the relation:
\begin{eqnarray}
(\gamma_5)_M=(\gamma_5)_E
\end{eqnarray}
From eqs. \eqref{eq-a2} and \eqref{eq-a4}, we have
\begin{eqnarray}
(\slashed{p})_M=(\gamma^0p^0-\gamma^kp^k)_M=i(\gamma^0p^0+\gamma^kp^k)_E=i(\slashed{p})_E
\end{eqnarray}
Likewise, $(\slashed{\partial})_M=i(\slashed{\partial})_E$,
$(\slashed{v})_M=i(\slashed{v})_E$, and
$(\slashed{a})_M=i(\slashed{a})_E$. And we let the other quantities
$\psi$, $\bar{\psi}$, and $m$ keep unchanged in these two spaces.
Thus, finally we obtain
\begin{eqnarray}
&&\exp\bigg[i\int
d^4x\bar{\psi}(i\slashed{\partial}+\slashed{v}+\slashed{a}\gamma_5-s-m)\psi\bigg]_M\nonumber\\
&&=\exp\bigg[-\int
d^4x\bar{\psi}(\slashed{\partial}-i\slashed{v}-i\slashed{a}\gamma_5-s+m)\psi\bigg]_E
\end{eqnarray}

\section{Necessary Integral Formulas}
In this appendix we list the integral formulas needed in the text as
follows:
\begin{eqnarray}
&&\lim_{\Lambda\rightarrow\infty}\int_{\frac{1}{\Lambda^2}}^\infty\frac{d\tau}{\tau}e^{-\tau
m^2}=\lim_{\Lambda\rightarrow\infty}(-\gamma-\ln\frac{m^2}{\Lambda^2})\label{eq-b1}\\
&&\lim_{\Lambda\rightarrow\infty}\int_{\frac{1}{\Lambda^2}}^\infty d\tau e^{-\tau m^2}
=\frac{1}{m^2}\\
&&\lim_{\Lambda\rightarrow\infty}\int_{\frac{1}{\Lambda^2}}^\infty
d\tau\tau e^{-\tau m^2}=\frac{1}{m^4}\label{eq-b3}
\end{eqnarray}
where $\gamma$ is the Euler-Mascheroni constant,
$\gamma\approx0.5772$.

\vspace{0.6cm}

{\bf Acknowledgments} \\

This work is supported by the Specialized Research Fund for the
Doctoral Program of Higher Education (SRFDP) under Grant No.
200805581030, and Sun Yet-Sen University Science Foundation.


\end{document}